\documentstyle[12pt,psfig]{article}
\catcode`\@=11
\@addtoreset{equation}{section}

\font\tenrm=cmr10
\font\tenit=cmti10

\def\rly#1{\mathrel{\raise.3ex\hbox{$#1$\kern-.75em\lower1ex\hbox{$\sim$}}}}
\def\lsim{\rly<}

\def\shat{\hat s}
\def\that{\hat t}
\newcommand{\k}{{\bf k}}
\newcommand{\p}{{\cal P}}
\newcommand{\K}{{\bf K}}
\newcommand{\rrr}{{\bf r}}
\newcommand{\R}{{\bf R}}
\newcommand{\ktc}{{\bf k}_c}
\renewcommand{\k}{{\bf k}}
\newcommand{\kta}{{\bf k}_a}
\newcommand{\ktb}{{\bf k}_b}
\newcommand{\vone}{{\bf p}_1}
\newcommand{\vtwo}{{\bf p}_2}
\newcommand{\vb}{{\bf k}}
\newcommand{\V}{{\bf K}}
\newcommand{\beq}{\begin{equation}}
\newcommand{\beqn} {\begin{eqnarray}}
\newcommand{\eeq}{\end{equation}}
\newcommand{\eeqn} {\end{eqnarray}}

\newcommand{\amp}{{\cal A}}
\newcommand{\di}{d^2}
\newcommand{\e}{{\cal E}}
\def\rgap{\hbox{$\Delta y$}}
\def\half{{\textstyle{1\over2}}}
\def\arctanh{{\rm arctanh}}
\def\tcm{\hbox{$\tilde{\cal M}$}}

\begin{document}
\rightline{UdeM-TH-95}
\rightline{McGill/95-49}
\rightline{Brown-HET-1023}
\rightline{hep-ph/9511252}
\bigskip
\rightline{November 1995}
\bigskip
\bigskip
\centerline{\bf PARTICLE PRODUCTION }
\centerline{\bf IN A HADRON COLLIDER RAPIDITY GAP:}
\centerline{\bf THE HIGGS CASE}
\bigskip
\centerline{ JEAN-REN\'E CUDELL\footnote{permanent address from Dec. 1995:
Universit\'e de Li\`ege, Physics Dept, B\^at. B-5, Sart Tilman,
B-4000 Li\`ege, Belgium}}
\centerline{\tenit Dept. of Physics,
McGill University, Montr\'eal QC, Canada H3A 2T8}
\centerline{\tenit
and Dept. of Physics, Brown University\footnote{Supported
in part by USDOE contract DE-FG02-91ER 40688-Task A.}
, Providence RI 02906, U.S.A.}
\centerline{\small cudell@hep.physics.mcgill.ca}
\medskip
\centerline{and}
\medskip
\centerline{OSCAR F. HERN\'ANDEZ }
\centerline{\tenit Labo. de Physique Nucl\'eaire,
Universit\'e de Montr\'eal, Montr\'eal QC Canada H3C 3J7}
\centerline{\small oscarh@lps.umontreal.ca}
\bigskip\bigskip\bigskip\bigskip\bigskip\bigskip
\centerline{\bf ABSTRACT}
\medskip
\begin{quote}

Production of rare particles within rapidity gaps has been propo\-sed
as a back\-ground-free signal for the detection of new physics at
hadron colliders.  No complete formalism accounts for such processes
yet. We study a simple lowest-order QCD model for their description.
Concentrating on Higgs production, we show that the calculation of the
cross section $pp\rightarrow pp H$ can be embedded into existing
models which successfully account for diffractive data. We extend
those models to take into account single and double diffractive cross
sections $pp\rightarrow HX_1 X_2$ with a gap between the fragments
$X_1$ and $X_2$. Using conservative scenarios, we evaluate the
uncertainties in our calculation, and study the dependence of the
cross section on the gap width. We predict that Higgs production
within a gap of 4 units of rapidity is about 0.3 pb for a 100 GeV
Higgs at the Tevatron, and almost 2 pb for a 400 GeV Higgs within a
gap of 6 units at the LHC with $\sqrt{s}$=14~TeV. \end{quote} \newpage

\section{Rapidity gaps as a detection tool}

At hadron colliders, rare high-$Q^2$ processes can be produced in the
usual hard collision picture, where two partons, one from each
nucleon, collide head-on and annihilate into the rare particle one is
trying to produce. The big drawback of such a production mechanism is
that hadrons cannot simply loose their partons, since the annihilation
of two partons leaves the initial nucleons in a coloured state. This
means that they will not fragment independently, but rather that a
colour string will link them. When they pull apart, the spectator
partons will break the string.  Hadronic matter will then populate
most of the central phase space. However in roughly 10\% of the hard
scattering events at HERA, further interactions neutralize the colour
of the proton fragments, and lead to large rapidity
gaps~\cite{HERAgaps}.

If one manages to produce a rare particle {\it without changing the
colour} of the nucleons, then very little hadronic activity should be
present in the event. Of course, both nucleons will still fragment, as
the production of a heavy particle imbalances the kinematics of their
partons, but the two fragmentations will be independent, as they are
not correlated by colour strings. Hence the produced hadrons will
follow the direction of the initial nucleons and there will be a large
rapidity gap with no hadronic remnants.  The rare particle will often
be produced within the gap, and hence its decay should be essentially
background-free. Bjorken~\cite{bj} has suggested using such rapidity
gaps as a means of detecting new physics.

Colour-singlet exchange between protons is in fact a very common
event, and it has been known and observed for a long time. Indeed, the
simplest kind of event with a rapidity gap is the elastic scattering
$pp\rightarrow pp$, which accounts for about 20\% of the total cross
section at the Tevatron~\cite{CDF,E710}. Another 20\% of the cross
section is due to single diffractive scattering $pp\rightarrow pX$,
and to double diffractive scattering $pp\rightarrow X_1 X_2$, both
processes leading to a large rapidity gap between produced hadrons.
Hence the study of rare particle production in a rapidity gap can be
seen as the inclusion of a high-mass component into soft, low-$Q^2$
diffractive physics.

Several attempts have been made to describe this process within a
structure function formalism, following Ingelman and
Schlein~\cite{IS}, and the demonstration by UA8 that hard diffractive
scattering does exist~\cite{UA8}.  These works use a pomeron structure
function, and treat the pomeron in a way similar to a photon. The
major drawback of this approach is that we do not know what the
pomeron is made of, and whether the concept of structure function
holds for it. Similarly, the pomeron structure function is not well
measured, even if it exists, and its flux factor is also subject to
question~\cite{goulianos}.

The cleanest estimate of Higgs production is presumably that due to
Bialas and Landshoff~\cite{BL}, where they perform a two-gluon
exchange calculation.  However, their calculation suffers from a few
drawbacks:\\ $\bullet$ They rely on the
Landshoff-Nachtmann~\cite{landshoffnachtmann} model of the gluon
propagator, which is taken to be a falling exponential. Although this
model has reasonable phenomenological support, it is not adapted to
higher-order calculations;\\ $\bullet$ In the case of diffractive
scattering, {\it i.e.} when the proton breaks up, they rely heavily on
Regge theory, as their cross section converges only because of the
non-zero Regge slope;\\ $\bullet$ They totally neglect exchanges
involving several quarks in each proton ;\\ $\bullet$ They neglect the
effects of the longitudinal parton kinematics, which is essential for
heavy particle production.

In this paper we remedy all the above problems: we introduce new form
factors, which describe diffractive scattering, and explain how the
longitudinal kinematics comes into play. We keep the exact kinematics
of the problem, and show that the cross section are IR finite even in
perturbative QCD. We then estimate the size of the cross section, and
obtain a result surprisingly close to that of Bialas and Landshoff.

It is well known that lowest-order QCD produces surprisingly good
estimates of diffractive processes.  In the present case, we tune the
form factors so that they reproduce the measured elastic and
diffractive cross sections. We then embed Higgs production within soft
cross sections.

Our first task will then be (in Section~2) to review the calculation
of the elastic cross sections~\cite{soper,cudellnguyen}, and to extend
it to single- and double- diffractive processes.  We will then (in
Section~3) lay out the formalism which embeds high-mass particle
production in diffractive physics.  Although the method is quite
general and can be used for a variety of processes, we shall
concentrate in Section~4 on Higgs production.  We show that our
treatment leads to a cross section which is about 20\% higher than the
estimate of Bialas and Landshoff, and that the details of the infrared
region do not play an overwhelming role in the determination of the
total rate. We also find that the evaluation of the elastic cross
section depends very little on the details of the model.  We then
evaluate the inelastic diffractive contribution $pp\rightarrow X_1X_2
H$, which turns out to be a factor $\sim 3$ larger than the elastic
one.  The final section contains a summary of our results and
conclusions.

\section{QCD models of diffractive physics}
Colour-singlet interactions between nucleons is an old topic.  Regge
theory attributes these interactions to the exchange of mesons,
grouped into Regge trajectories according to their quantum numbers.
In the high-$s$ limit, one expects the hadronic amplitude to be a sum
of simple poles, each pole corresponding to the exchange of the
particles lying on a given Regge trajectory:
\begin{equation}
{\cal A}(s,t)=\sum_i \beta_i^2(t)\ s^{\alpha_i(t)} \xi(\alpha_i(t))
 \label{hadronic}
\end{equation}
with $\xi$ the Regge signature factor.  ${\cal A}(s,t)$ is normalised so that
the elastic cross section is given by
$d\sigma_{el}/dt=|{\cal A}|^2/16\pi^2$, which
through the use of the optical theorem gives a total cross section
$\sigma_{tot}
=Im\left({\cal A}(s,t=0)\right)$.

The leading meson trajectories are the
(degenerate) trajectories of the $\rho$ and $f$ mesons, clearly present
 when plotting the meson states in a $J$ vs $M^2$ plane.
When continued to negative values of $M^2$, these trajectories are responsible
for the fall-off of the total cross section at small $s$, $\sigma_{tot}
\sim 1/\sqrt{s}$.
At higher energies, the cross sections grow, and the most natural assumption is
that another term in Eq.~(\ref{hadronic}) is responsible for
that rise although there is no observed trajectory to guide us.
Hence one new Regge trajectory
was invented, the pomeron, which is whatever makes the total cross section
rise at high energy (at 1800 GeV, more than 99\% of the total cross section is
due to pomeron exchange). In modern day QCD,
non-mesonic  exchanges are attributed to multi-gluon exchanges, which
naturally  produce a rising elastic cross section in perturbation theory.

Several other properties of high-$s$ colour-singlet exchanges can be
inferred from data~\cite{LD}:\\
{$\bullet$} The pomeron trajectory of Eq.~(\ref{hadronic})
is well fitted from $\sqrt{s}=5$ GeV to $1800$ GeV, and
{}from $t=0$ to $-1$ GeV$^2$ to the form
\begin{equation}
\alpha_P(t)=1.08+0.25 t
\label{pomtraj}
\end{equation}
{$\bullet$} The pomeron
coupling to protons, $\beta_P(t)$, is well approximated by the coupling of
photons, {\it i.e.} by the Dirac elastic form factor:
\begin{equation}
F_1(t)={(3.53-2.79 t)\over(3.53-t)\ (1-t/0.71)^2}\label{F1}
\end{equation}
\noindent{$\bullet$} Colour-singlet exchanges are factorizable:
the ratio of the differential cross section  for $p A\rightarrow XA$ to
that for $pA\rightarrow pA$ does not depend on $ A$~\cite{Goulianos}. Another
related property is the quark counting rule: hadronic cross sections are
 proportional to the number of valence quarks contained
in the hadron. This is tested not only in $\pi p$ vs $pp$ cross sections, but
also in cross sections involving strange quarks,
{\it e.g.} $2\sigma_{tot}(\Omega^- p)-\sigma_{tot}(\Sigma^-p)$ can be
 predicted from $\sigma_{tot}(pp)$, $\sigma_{tot}(\pi p)$ and
$\sigma_{tot}(Kp)$~\cite{Quarkcounting}.

\subsection{QCD models for elastic and total cross sections}
\begin{figure}
\centerline{\psfig{figure=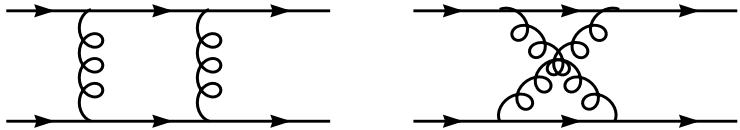,width=10cm}}
{\caption{\tenrm \baselineskip=12pt The lowest order Feynman diagrams
for $pp$ colour-singlet exchange}
\label{fig2}}
\end{figure}

The simplest QCD colour-singlet exchange consists of two gluons~\cite{Low}.
and is shown in Fig.~\ref{fig2}.
If we exchange these gluons between quarks, we obtain the following
(IR divergent) amplitude~\cite{cudellnguyen}:
\begin{equation}
{\cal A}_q(\shat,\that)= {8i\over 9}\ \alpha_S^2 s  \int d^2\kta d^2\ktb
{d{\cal A}_q\over d^2\kta d^2\ktb}\label{qqscatt2}
\end{equation}
with
\begin{equation}
{d{\cal A}_q\over d^2\kta d^2\ktb}=
\delta^{(2)}({\bf\Delta} - \kta - \ktb) {1\over ({\kta}^2-\sigma_a)}
\times{1\over
({\ktb}^2-\sigma_b)}\label{qqscatt3}\end{equation}
Here bold face variables represent transverse momenta. The transverse
momenta $\kta$ and $\ktb$ are the components of the gluon momenta
transverse to the direction of the quarks and ${\bf\Delta}$ is the
total momentum exchanged by the quarks (${\bf\Delta}^2=t$).
We have introduced two gluon squared masses $\sigma_a$ and
$\sigma_b$.

The quark counting rule suggests that
we describe a proton as made of its 3 valence quarks, and that the sea
quarks do not contribute much to these processes (or that they are generated
by higher orders). One can then show that quark-quark scattering can be
simply embedded into the proton~\cite{cudellnguyen,soper}.  The elastic
amplitude for two-gluon exchange between two protons has been shown to be:
\begin{equation}
\amp_2= {8 i}\alpha_S^2 \shat \int \di\kta
\di\ktb {\di\amp^q_2\over \di\kta
\di\ktb}
[{\cal E}_1(\kta+\ktb)-{\cal E}_2(\kta,\ktb)]^2
\label{hhscatt}\end{equation}
${\cal E}_1$ and ${\cal E}_2$ are two of the three form factors that
can occur in the valence quark description of a proton.  Each of these
form factors correspond to the situation where 1, 2 or 3 quarks get
hit by gluons.  One can write their expression in terms of the proton
wavefunction $\psi$:
\beqn {\cal E}_1({\bf k})&=&\int d{\cal M} |\psi(\beta_j,\rrr_j)|^2
e^{i \k.\rrr_k}\nonumber\\
      {\cal E}_2(\k_a,\k_b)&=&\int d{\cal M} |\psi(\beta_j,\rrr_j)|^2
e^{i \k_a.\rrr_k+i \k_b.\rrr_l}\nonumber\\
      {\cal E}_3(\k_a,\k_b,\k_c)&=&\int d{\cal M}\ |\psi(\beta_j,\rrr_j)|^2
e^{i \k_a.\rrr_k+i\k_b.\rrr_l+i\k_c.\rrr_m}\label{f}
\label{formone}
\eeqn
where $\beta_j$ is the fraction of longitudinal momentum (similar to
Bjorken $x$),
and $\rrr_j$ is the transverse position of quark $j$.
The natural integration measure $d{\cal M}$ is defined as:
\beq d{\cal M}=[\prod_{j=1,n_q} d\beta_j d^2\rrr_j ]\delta^{(2)}
(\sum_{j}\beta_j \rrr_j)
\delta(\sum_{j}\beta_j-1) \label{measure}\eeq
The first delta function defines the center of momentum of the hadron, whereas
the second one enforces longitudinal-momentum conservation.
Assuming that hadrons are made of valence quarks only, we normalise the
wavefunction according to:
\beq \e_1(0)=\int d{\cal M}|\psi(\beta_j,\rrr_j)|^2 =1 \label{norm}\eeq

These expressions become useful once one realises that the form factor
${\cal E}_1$ also occurs in the elastic $\gamma^* p$ cross section and
is none other than the Dirac elastic form factor in Eq.~(\ref{F1}).
Hence one of the form factors is determined. Furthermore, ${\cal E}_2$
and ${\cal E}_3$ are related to ${\cal E}_1$ by the following
properties:
\beqn
\e_2(\kta,0)&=&\e_1(\kta)\nonumber\\
\e_3(\kta,\ktb,0)&=&\e_2(\kta,\ktb)\label{IR}
\eeqn
These properties ensure the IR finiteness of Eq.~(\ref{hhscatt}).
One can either calculate these form factors using a model for the proton wave
function~\cite{cudellnguyen}, or use a simple parametrisation, which takes
into account both the infrared and the symmetry
properties of the form factors: one simply needs to make an ansatz for
$\e_3$ and obtain $\e_2$ from Eq.~(\ref{IR}).  We take
\beq
\e_3(\kta,\ktb,\ktc)=\e_1\left(\kta^2+\ktb^2+\ktc^2-c(\kta\cdot\ktb+\kta\cdot
\ktc+\ktb\cdot\ktc)\right)
\label{formtwo}
\end{equation}
where $c$ is an arbritrary number
which can
be shown to be of order one in  the proton case~\cite{cudellnguyen}.

We have written the form factors as functions of the gluon
momenta to manifestly show that Eqs.~(10) hold as the gluon momenta go on
shell.
However the form factor comes from the proton wavefunction and
can only depend on quark variables.
Hence $\kta,~\ktb,~\ktc$ should be thought of as differences between initial-
and final-quark transverse momenta.

At this stage, one is in a situation to describe elastic scattering and hence,
through the use of the optical theorem, total cross sections. The model
contains two free parameters, $\alpha_S$ and $c$, which we can tune to
reproduce the data. One can then look at the
shape of the differential elastic cross section to check whether the model
makes sense. Unfortunately, it is
well-known~\cite{soper,cudellnguyen,levinryskin} that the shape of the cross
section comes out wrong: its logarithmic slope at the origin is infinite, and
its curvature is too big.

One may be tempted to look into higher order corrections for a
solution to this problem. Indeed, $\alpha_S$ is not small, and the
perturbative series contains terms of order $\alpha_S^{n+1} \log^n s$
which clearly can become big at large $s$. These terms can be resummed
using the BFKL techniques~\cite{BFKL}, but such formalism cannot
account for diffractive scattering at low momentum transfers, where
unfortunately most of the cross section is concentrated.  The most
obvious problem is that the rise of the cross section predicted by
leading-log-$s$ perturbative QCD is entirely different from that which
is observed in data, as its leading contribution to the hadronic
amplitude goes like $s^{1+2.65\alpha_s}$. Secondly, the slope of the
pomeron trajectory $\alpha'=0.25$~GeV$^{-2}$ introduces a scale of the
order of $ 1/\sqrt{\alpha'}\approx 2$ GeV, which comes in the
description of the $t$-dependence of the hadronic amplitude. No such
scale is present in perturbative QCD, hence the differential elastic
cross section retains its wrong shape.  Finally, the amplitude is not
factorizable: this is in fact a consequence of the infrared finiteness
of the answer. Quark-quark scattering via gluon exchange diverges for
massless gluons. Nevertheless, hadron-hadron scattering is infrared
finite, as the colour of the hadron gets averaged for very long
wavelength gluons. Hence there is a contribution that comes from the
diagrams where gluons are exchanged between different quarks in the
hadron. These diagrams feel the hadronic wavefunction, and hence their
contribution depends on the target.

Even the lowest-order two-gluon exchange cross section fails to
reproduce the factorizability of the hadronic amplitude.  This
prompted Landshoff and Nachtmann~\cite{landshoffnachtmann} to
postulate that gluons have an intrinsic propagation length $1/\mu_0$
smaller than typical hadronic sizes. The propagator hence becomes:
\begin{equation}
D(k^2)=d(k^2/\mu_0^2)/\mu_0^2\label{npglue}
\end{equation}
with $\lim_{x\rightarrow 0} xd(x)=0$.  In that case, the diagrams in
which the gluons couple to different quarks are suppressed by a factor
$\sim (25\ GeV^{-2})\mu_0^2$ with respect to those in which the gluons
couple to the same quark line, and factorisation can be restored, for
$\mu_0\sim 1$ GeV.

This kind of picture has recently been questioned on theoretical
grounds~\cite{Buttner}, but it seems to remain the only one available
which reproduces the observed properties of the pomeron.  In the
following, rather than worrying about the ultimate nature of the
pomeron, we shall adopt the following pragmatic approach: we are after
rates for the production of rare particles produced in a rapidity gap.
This is very similar to single-diffractive processes and elastic
processes, except that we need to insert a rare particle production
vertex in the diagram. The best phenomenological description of such
processes is two-gluon exchange, multiplied by a pomeron s-dependence
given by Eq.~(\ref{hadronic}).  We shall first set up the lowest-order
perturbative calculation of Higgs production via two-gluon exchange.
We shall then evaluate the importance of the IR region by using a
constituent-gluon propagator
which matches to the perturbative one at high $k^2$~\cite{cudellross}.
The model hence contains several
parameters: $\alpha_S$, $c$ and a non perturbative scale $\mu_0$. We
shall tune these parameters to the highest-energy data available,
reproducing the total, elastic and diffractive cross sections at the
Tevatron. Hence our calculation of Higgs production at the LHC will be
parameter-free.

Note that this procedure also factors in the gap survival. Indeed,
diffractive cross sections surely produce a gap, and the final-state
multiple interactions which may spoil it are taken into account by our
tuning of the parameters. As the kinematics of the process we are
considering is rather similar to diffractive scattering, we expect
that the choice of parameters which accounts for gaps in diffractive
scattering will also lead to gaps in Higgs production.

\subsection{A QCD model for inelastic diffractive cross sections}
Before embarking into the afore-mentioned fit, we need to develop the
formalism needed to describe inelastic diffractive cross sections,
{\it i.e.} those in which a proton gets hit by a colour singlet, but
nevertheless fragments. The main ingredient needed is a new form
factor for the proton.

Hence we want to describe the process $p\rightarrow X$ instead of
$p\rightarrow p$. The simplest way to do this is to square the
amplitude, and to recognise that the cross section, including all the
interference terms in the final state, is similar to the elastic
amplitude resulting from 4-gluon exchange, with gluons arranged in two
singlets.

To calculate the corresponding form factor, one has to notice that
one-photon exchange leads to an ${\cal E}_1$ form factor. The
two-gluon form factor can then be built by squaring the one-photon
exchange diagram, and removing the intermediate proton. This can lead
to an ${\cal E}_1$ form factor, if the same quark is hit in the two
interfering diagrams, or to an ${\cal E}_2$ form factor, if different
quarks are hit. Similarly, an ${\cal E}_2$ convoluted with an ${\cal
E}_1$ will give either an ${\cal E}_2$, or an ${\cal E}_3$. In the
case of the calculation of the elastic form factor, there is an extra
complication coming from the colour algebra. However, in the case at
hand, the colour algebra is trivial, because we are exchanging
colour-singlets, and hence the colour factor is factorised between the
two sides of the diagram.  Therefore, after we recognise that the
elastic form factors are given by 9 terms, three ${\cal E}_1$ and six
$ {\cal E}_2/2$, we can simply write down the 81 possible
interferences, and decide by inspection which form factor will
describe them.

More formally, one first needs to assume that in diffractive
scattering one goes through an intermediate state, call it $Y$,
described by a wave-function $\psi_Y$. The only change in the
preceeding formulae~(\ref{formone}) then occurs in the substitution
\break $|\psi(\beta_j,r_j)|^2 \rightarrow \psi^*_p(\beta_j,r_j)
\psi_Y(\beta_j,r_j)$. This means that we are not going to describe
intermediate states which have a mass very different from that of the
proton, as their production necessitates a change in the longitudinal
momentum. Hence the cross section we are calculating represents the
bulk of diffractive scattering, but does not reproduce the high-mass
tail. This tail presumably comes from higher-order diagrams, be they
3-pomeron vertices or gluon radiation\cite{ryskin,Goulianos}.

Now, in elastic scattering, the form factor describing $p\rightarrow
p$ is $3({\cal E}_1(t)-{\cal E}_2(t))$.  This comes from $9\times 2$
diagrams similar to those of Fig.~{\ref{fig2}}, as each gluon can be
attached to 3 different quarks. The form factor describing
$p\rightarrow Y$ is the similar, with {\it e.g.} ${\cal E}_1$ given by
\beq
{\cal E}_1(\k)=\int d{\cal M}
 \psi^*_p(\beta_j,r_j) \psi_Y(\beta_j,r_j) e^{i \k.\rrr _k}
\label{newcale1}
\eeq

\begin{figure}
\centerline{\psfig{figure=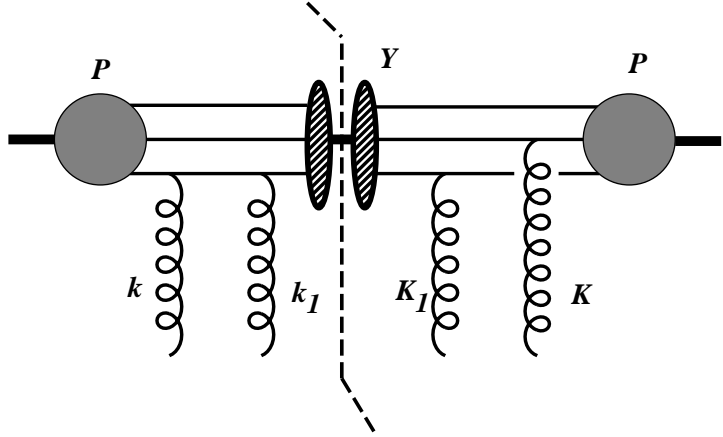,width=10cm}}
{\caption{\tenrm \baselineskip=12pt One of the 81
interference terms contributing to the inelastic diffractive form factor}
\label{extrafig}}
\end{figure}

Squaring the  amplitude, we obtain
the square of these form factors. Let us for instance concentrate on the
${\cal  E}_1 \times {\cal E}_2$ term shown in Fig.~\ref{extrafig}.
It corresponds to the expression:
\beqn \sum_Y {\cal E}_1(\k+\k _1){\cal E}_2(\K _1,\K)&=&\sum_Y\int
[\prod_{j=1,n_q} d\beta_j d\rrr _j ]\delta^{(2)}
(\sum_{j}\beta_j \rrr _j)
\delta(\sum_{j}\beta_j-1) \nonumber\\
&&\psi^*_p(\beta_j,\rrr _j)  \psi_Y(\beta_j,\rrr _j)  e^{i (\k+\k _1).\rrr _1}
\nonumber\\
&\times&[\prod_{l=1,n_q} dB_l d\R _l ]\delta^{(2)}
(\sum_{l}B_l \R _l)
\delta(\sum_{l}B_l-1) \nonumber\\
&&\psi_Y^*(B_l,\R _l)  \psi_p(B_l,\R _l)  e^{i \K _1.\R _1+\K .\R _2}
\label{f1f2}\eeqn
The completeness relation for the wavefunctions reads:
\beqn f(\beta_l,\rrr _l) = \sum_Y \int
[\prod_{j=1,n_q} dB_j d\R _j ]\delta^{(2)}
(\sum_{j}B_j \R _j)
\delta(\sum_{j}B_j-1) \nonumber\\
\psi_Y(\beta_l,\rrr _l)  \psi_Y^*(B_j,\R _j) f(B_j,\R _j)
\eeqn
Hence formula (\ref{f1f2}) becomes:
\beqn \sum_Y {\cal E}_1(\k+\k _1){\cal E}_2(\K _1,\K)&=&\int d{\cal M}
e^{i (\k+\k _1).\rrr _1}|\psi_p(b_l,\rrr _l)|^2  e^{i \K _1.\rrr _1+\K .\rrr
_2}
\nonumber\\
&=& {\cal E}_2(\k +\k _1+\K _1,\K)\eeqn
All the other 80 interference terms can be worked out similarly, and
 we get the form factor in the inclusive case from
the square of the amplitudes.

Noting that the four momenta entering the form factor have to sum to zero (as
we are squaring an amplitude, the initial and the final states must be
identical), we use the short-hand notation:
$\e_2({\bf l})\equiv\e_2({\bf l},-l)$, $\e_3({\bf l},{\bf L})
\equiv\e_3({\bf l},{\bf L},-{\bf l}-{\bf L})$.
We then obtain the following:
\beqn 3&{\cal F}&(\k ,\k _1,\K _1,\K ) = 3\ [1-\e_2(\k )-
\e_2(\k _1)-\e_2(\K _1)
-\e_2(\K )\nonumber\\
&+&2\e_2(\k +\k _1)+{1\over 2}\e_2(\k +\K)+{1\over 2}\e_2(\k+\K _1)
\label{formdiff}\\
&-&\e_3(\k ,\k _1)-\e_3(\K ,\K _1)+{1\over
2}\e_3(\k _1,\K _1)+{1\over 2}\e_3(\k ,\K _1)+{1\over 2}\e_3(\k ,\K )+{1\over
2}\e_3(\K ,\k _1)]\nonumber\eeqn
Note that we recover the same IR behaviour of the square form factor as we
had for its elastic components:
$ {\cal F}(\k _1,\k _2,\k _3,\k _4)\rightarrow 0$ for any
$\k _i^2\rightarrow 0$. This
insures the infrared finiteness of the answer.

The diffractive cross section then has the same form as the square of the
elastic amplitude, except that the form factor $81 [{\cal E}_1-{\cal E}_2]^4$
gets replaced by $9{\cal F}^2$.

\subsection{Best values of parameters}

We are now in a position to fix the parameters of the model.
As we need to extrapolate to the LHC, we use Tevatron data. The fits we obtain
are good to about $10\%$ in the perturbative case. Slightly better results
are obtained if we
smooth the infrared region of the propagator, using a constituent-gluon
propagator, as explained at the end of Section 2.1.

\begin{center}
{\bf Table 1:} Diffractive data\end{center}

\begin{center}
\begin{tabular}{|c|c|c|c|} \hline
 experiment& $\sigma_{tot}$ (mb) &$\sigma_{el}$ (mb) &$2\sigma_{sd}$ (mb)\\
\hline
 CDF~\cite{CDF}& $80.03\pm 2.24$&$19.7\pm 0.85$&$9.46\pm 0.44$\\ \hline
 E710~\cite{E710} &$72.2 \pm 2.7$&$16.6\pm 0.7 $&$9.37\pm 2.9$  \\ \hline
\end{tabular}
\end{center}

The E710 data are well reproduced by the following set of parameters:
\begin{center}
{\bf Table 2:} Parameters reproducing the data of Table 1\end{center}
\begin{center}
\begin{tabular}{|c|c|c|c|} \hline
propagator& $\alpha_S$ &$c$ &$\mu_0$ (GeV)\\ \hline
perturbative & 0.88 &0.61 &-\\ \hline
constituent~\cite{cudellross}&        1.52 &0.11 & 1.9\\ \hline
\end{tabular}
\end{center}
Note that in the case of the constituent gauge propagator,  the running of
the coupling is included in the gluon propagator, and that in this case
the value of $\alpha_S$ is that at the renormalisation point $\mu_0$.

\begin{figure}
\centerline{\psfig{figure=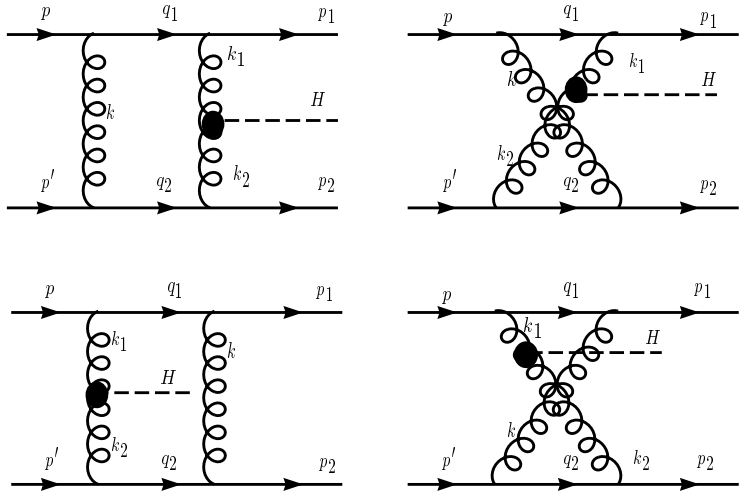,width=10cm}}
{\caption{\tenrm \baselineskip=12pt The Feynman diagrams contributing to H
boson production in a rapidity gap. }
\label{fig1}}
\end{figure}

\section{ Heavy Particle Production}

The purpose of this work is to estimate the cross section for the production
of a heavy particle H
within the central region, hence the heavy particle must be
produced within the gluon exchange, and not as a result of the proton
fragmentation.
One can describe the production mechanism via an effective
gluon-gluon-H vertex:
\beq
W_{\mu\nu}=\delta^{ab} (k_1\cdot k_2 g^{\mu\nu}-k_1^\mu k_2^\nu){W_1 \over
m_H^2}
+ (k_1^2 k_2^2 g^{\mu\nu} + k_1^\mu k_2^\nu k_1 \cdot k_2
    - k_1^\mu k_1^\nu k_2^2 - k_2^\mu k_2^\nu k_1^2){W_2 \over m_H^4}
\label{effvert}
\eeq
and the specialisation to a specific particle is done via the calculation of
$W_1$ and $W_2$.

Once this vertex is known, we must embed it into the two-gluon exchange
graphs. One can couple the vertex to either gluon, and there are two two-gluon
exchange graphs, hence in principle we have to calculate four
diagrams. However, as noted in Ref.~\cite{BL}, the sum of the four
diagrams in Fig.~\ref{fig1} is equal to the $s$-channel discontinuity
of the first one.  Thus we can obtain the quark-level H production
cross section by calculating the imaginary part of the diagram in
Fig.~\ref{fig4}.

Heavy particle production forces the gluons to carry a non-vanishing
longitudinal momentum, hence the kinematics is not purely transverse anymore.
We shall have to modify slightly the expressions for the form factors to take
this into account, and these form factors will then allow us to go from
the process
$qq\rightarrow Hqq$ to $pp\rightarrow Hpp$, or $pp\rightarrow HX_1X_2$.

\subsection{Kinematics}

The expression for the quark-level cross section is given by:
\beq
d\hat\sigma={(2/9)^2\over 8 \shat} |{\cal M}|^2
\label{cross}
\eeq
with the $2/9$ the colour factor for two-gluon exchange.

\begin{figure}
\centerline{\psfig{figure=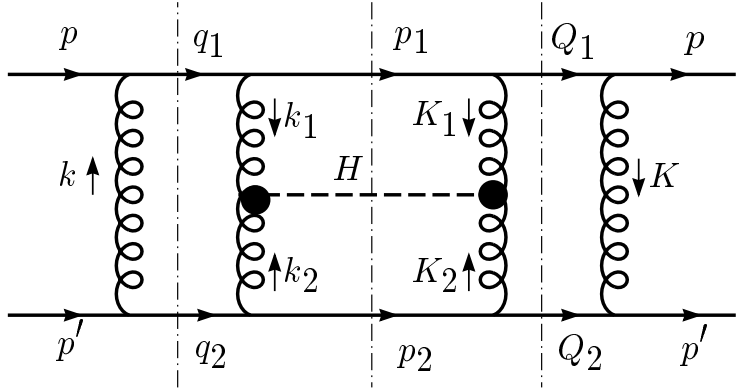,width=10cm}}
{\caption{\tenrm \baselineskip=12pt The square of the
Feynman diagram contributing to H production in a rapidity
gap. The imaginary part of this diagrams gives the total cross section
for the process in Fig.~\protect\ref{fig1}.  The dash-dotted line shows the
cuts used in obtaining the imaginary part. }
\label{fig4}}
\end{figure}

The contribution to the total cross section of the process depicted in
Fig~\ref{fig1} is arrived at by calculating the imaginary part of the
diagram in Fig~\ref{fig4} via cutting rules.  In the notation of
Fig.~\ref{fig4}, the square of the imaginary part of the amplitude
given by:
\beqn
|{\cal M}|^2
&=&{\alpha_S^4\over 2\pi^5}\int{d^4k} {d^4K} {d^4p_1} {d^4p_2}
\delta_+(p_1^2)
\delta_+(p_2^2)
\delta_+(q_1^2)
\delta_+(q_2^2)
\delta_+(Q_1^2)
\delta_+(Q_2^2)
\nonumber\\
& &\delta_+(H^2-M_H^2){\tilde{\cal M}}
\label{imaginamp}
\eeqn
where ${\tilde {\cal M}}$ is the squared differential amplitude given by
Feynman rules
(see the next section) and $\delta_+(k^2)=\delta(k^2)\theta(k_0)$.

Let us rewrite the momenta in terms of Sudakov variables:
\beqn
k&=&{\bar x\over \shat} p + {\bar y\over \shat} p' + \k \\
p_1&=&x_1 p +{\bar y_1\over \shat} p' +\vone \\
p_2&=&{\bar x_2\over \shat} p +{y_2} p' +\vtwo
\label{sudakov}
\eeqn
with $\vb ,\vone,\vtwo$ transverse to $p$ and
$p'$. \\
We assume that the off-shellness of the incoming quarks can be neglected,
so that
\beq
p.p=0, p.p'=\shat/2, p'.p'=0
\eeq
One can then solve for the $\delta$ functions putting the quarks on-shell:
\beqn
\delta(q_1^2) &=& \delta(\bar y+\vb ^2) \\
\delta(q_2^2) &=& \delta(\bar x-\vb ^2) \\
\delta(p_1^2) &=& {1\over x_1}\delta(\bar
y_1+{\vone ^2\over x_1}) \\
\delta(p_2^2) &=& {1\over y_2}\delta(\bar
x_2+{\vtwo^2\over y_2})
\label{deltas}
\eeqn
Treating the $\delta_+(Q_1^2)\delta_+(Q_2^2)$ in a similar manner, we
arrive at:
\beqn
d^4 k\ d^4 K\ \delta(q_1^2)\delta(q_2^2)\delta(Q_1^2)\delta(Q_2^2)&=&
{1\over 4\shat^2} d^2\vb  d^2\V  \\
d^4 p_1 d^4 p_2\ \delta(p_1^2)\delta(p_2^2)&=&
{1\over 4} {dx_1\over x_1} {dy_2\over y_2} d^2 \vone  d^2 \vtwo\\
\nonumber\label{newmeasure}
\eeqn

We now have to deal with the $\delta$-function putting the H
on-shell:
\beq
\delta((p_1+p_2-p-p')^2-M_H^2) = \delta( 2 \vone .\vtwo+ \shat (1-x_1)
(1-y_2) +{\vone ^2 \vtwo^2\over x_1 y_2}+
{\vone ^2\over x_1}+{\vtwo^2\over y_2}-M_H^2)
\label{higgsdelta}
\eeq
We will eventually use it to eliminate $x_1$ or $y_2$.

Hence the differential cross section becomes:
\beqn d\hat\sigma&=&{1\over 81\times 8 \shat}{\alpha_s^4\over (2\pi)^5 \shat^2
}{dx_1\over
x_1} {dy_2\over y_2} {d^2 \vone  d^2 \vtwo} d^2 \vb  d^2\V
\nonumber\\
&\times&\delta\left(\shat (1-x_1)(1-y_2) +(\vone +\vtwo)^2  +{\vone ^2\over
x_1}+{\vtwo^2\over
y_2}-M_H^2\right)\tilde{\cal M}\eeqn

The positivity of the energy of the on-shell lines dictates the integration
limits:
\beq
0<x_1,y_2<1 \ ,  \qquad
|\bar y+\bar x|<\shat \ , \qquad
|\bar Y+\bar X|<\shat \ .
\label{bounds}\eeq

All the dot products can then be expressed in terms of the transverse
kinematics and of the longitudinal momentum fractions $x_1$ and $y_2$. The
integration bounds (\ref{bounds}) do not guarantee the existence of a rapidity
gap, but simply that of a physical process.

Several conditions are necessary for the existence of a gap. First of
all, the final state must not be too far off-shell. If the incoming
proton has momentum ${\cal P}$, then the outgoing X cluster has
momentum ${\cal P}_{X}\equiv {\cal P}-p+p_1$. Its mass is given by
(write ${\cal P}=p/x_B\equiv\omega_B p$) $M_X^2 = |\vone |^2 {(\omega_B-1)/
x_1}$.  Similarly for the outgoing cluster $X'$ with momentum ${\cal
P}_{X'}\equiv {\cal P}'-p'+p_2$ we have $M_{X'}^2 =|\vtwo|^2
{(\omega'_B-1)/ y_2}$.

The rapidity gap \rgap\ between the two outgoing quarks is given by:
\beq
\rgap \equiv
\half\ln{p'.p_1 \over p.p_1}  - \half\ln{p'.p_2 \over p.p_2}
= \ln{x_1 y_2 \shat \over |\vone ||\vtwo|}\\
\label{rapgap}
\eeq
and that between the two hadronic clusters is:
\beq \rgap\equiv
\half\ln{{\cal P}'.{\cal P}_{X} \over {\cal P}.{\cal P}_{X}}  - \half\ln{{\cal
P}'.{\cal P}_{X'} \over {\cal P}.{\cal P}_{X'}}
={1\over 2} \log\left({s^2 x_1 y_2 (\omega_B + x_1 -1) (\omega'
_B +y_2-1)\over \omega_B^2 \omega^{\prime 2}_B \vone ^2 \vtwo^2}\right)\eeq

Setting aside the question of gap survival (see section 4.4),
a large rapidity gap will be present if the fractional
longitudinal momentum loss of the quarks $1-x_1$, $1-y_2$,
and the transverse momentum fraction, $|\vone ||\vtwo|/\shat$, are small.
The gluon propagators automatically provide us with small $|\vone
||\vtwo|/s$.

Whereas large gaps are produced for $x_1$ and $y_2$ near one, heavy particle
production requires that neither $x_1$ nor $y_2$ be too close to unity:
\beq
(1-x_1)(1-y_2)\hat s\approx M_H^2 \ .
\eeq
Since for valence quarks, $\hat s\approx s/9$, we arrive at the kinematic limit
\beq M_H\leq (1-x_1) \sqrt{s} /3
\label{massmax}
\eeq
We shall take $x_1>0.7$, since cross sections then become mainly
diffractive~\cite{Goulianos}.
Roughly speaking, particle production and
detection in a rapidity gap is limited to masses
less than an order of magnitude
below the center of mass energy.
Note that here $x_1$ is defined at the quark level, and that this cut is
similar to that used by Bialas and Landshoff.

\subsection{The s-channel discontinuity of the amplitude}

Let us now turn to the evaluation of $|{\cal M}|^2$
(see Eq.~(\ref{imaginamp})),
which is what couples to
the diffractive form factor.
Because of the flat
high-$s$ behaviour of the cross section, and because the exchange is
even under charge parity, the amplitude ${\cal M}$ is equal to its
$s$-channel discontinuity.

$\tilde{\cal M}$ is made of two traces, $T_1$ and $T_2$,
six gluon propagators, $D(k^2)$,
and two effective vertices from the Higgs loop, $\Phi_1^{N\Lambda}$ and
$\phi_1^{\nu\lambda}$. We thus have, (see Eq.~(\ref{imaginamp})):
\beq
\tcm=T_1 T_2 \Phi_1^{N\Lambda} \phi_1^{\nu\lambda}
D(k_1^2)D(k_2^2)D(k^2)D(K_1^2)D(K_2^2)D(K^2)
\label{tcmone}
\eeq
The traces are given by:
\beqn T_1&=& Tr(p.\gamma\ \gamma_\mu\ q_1.\gamma\ \gamma_\nu\ p_1.\gamma\
\gamma_{N}\ Q_1.\gamma\ \gamma_{M})\nonumber\\
 T_2&=& Tr(p'.\gamma\ \gamma_\mu\ q_2.\gamma\ \gamma_\lambda\ p_2.\gamma
\ \gamma_{\Lambda}\ Q_2.\gamma\ \gamma_{M})\eeqn
and the tensor structures in Eq.~(\ref{effvert}) associated with the
$ggH$ vertex are:
\beqn \Phi_1^{N\Lambda}&=&(K_1.K_2 g^{N\Lambda}
-K_2^{N} K_1^{\Lambda}) W_1/m_H^2 \nonumber\\
\phi_1^{\nu\lambda}&=&(k_1.k_2 g^{\nu\lambda}-k_2^\nu k_1^\lambda) W_1/m_H^2
\eeqn
Here, we have neglected the $W_2$ term, as it will be smaller in the special
instance of Higgs production (see section~4 and the Appendix).

The number of terms arising from the traces, after we perform a Sudakov
expansion of the momenta, is of the order of one million
and so we will keep only the leading order in the quark center of mass
energy $\shat$.
The final answer then takes the form:
\beq \tilde{\cal M}=\shat^4{\cal A}D(k_1^2)D(k_2^2)D(k^2)D(K_1^2)D(K_2^2)D(K^2)
\label{finalcalm}
\eeq
Here $D$ is the gluon propagator and $\cal A$
is a complicated function of the
transverse momenta, which includes the triangle vertex.
We do not quote the full expression for $\cal A$
as used in the program since that is not very illuminating. Instead
we present the first terms in a series expansion in
$1-x_1,\ 1-y_2$, which as discussed in the previous section, are required to
be small ($< 0.3$) to insure a large rapidity gap.
\beq {\cal A}=\left({\cal A}_0 +{\cal A}_1 (1-x_1)
+ {\cal A}_2 (1-y_2)\right) W_1
{W_1'}^* /M_H^4
\eeq
with:
\beqn {\cal A}_0 &=& 64\ (\vb .\vone  - \vb .\vtwo - \vb ^2)\ (\V .\vone  - \V
.\vtwo - \V ^2)\nonumber\\
{\cal A}_1 &=&  - 64\ [(\vb .\vone  - \vb ^2)\ (2\ \V .\vone  - \V .\vtwo - 2\
\V ^2) - (\V .\vone  -
\V ^2)\ \vb .\vtwo]\nonumber\\
{\cal A}_2 &=& 64\ [(\vb .\vtwo + \vb ^2)\ (\V .\vone  - 2\ \V .\vtwo - 2\ \V
^2) + (\V .\vtwo +
\V ^2)\ \vb .\vone ]\eeqn

\subsection{The form factor}

We have discussed elastic and diffractive form factors when all
gluon momenta are transverse.  We are however interested in producing
a Higgs particle in the rapidity gap. This requires that
a significant fraction of the proton longitudinal
momentum be carried by the gluons.  Our
previous expressions for the form factor needs to be adapted
to this situation.

\begin{figure}
\centerline{\psfig{figure=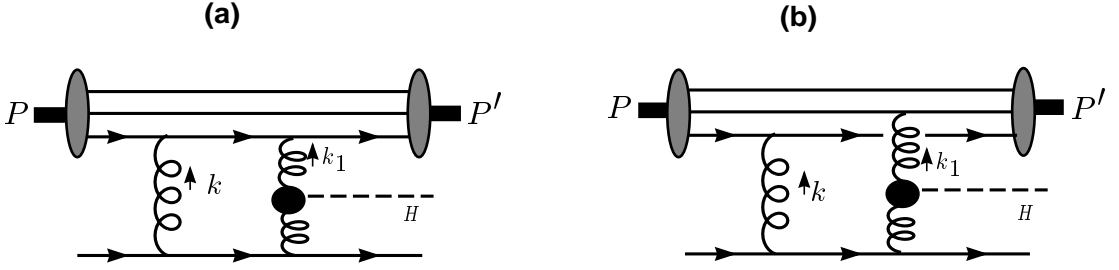,width=15cm}}
{\caption{\tenrm \baselineskip=12pt Valence quark description of proton
where one (a) or two (b) quarks get hit by a gluon. }
\label{fig5form}}
\end{figure}

Let us first consider
the process $pp\to ppH$, similar to the elastic cross section in
Eq.~(\ref{hhscatt}). This involves two form factors corresponding to
diagrams (\ref{fig5form}a) and (\ref{fig5form}b).
One starts with a proton of momentum $\cal P$ and ends with an on-shell
proton of momentum
\begin{equation}
{\cal P}_1=x_1 {\cal P}-{\vone ^2\over x_1 s}{\cal P}'+\vone
\end{equation}
${\cal P}'$ is the momentum of the proton with three momentum opposite
to that of ${\cal P}$.
What matters for the quark recombination, and hence for the form
factor, is the transverse kinematics in the ${\cal P}_1$ frame.
The total momentum transfer is $\Delta={\cal
P}_1-{\cal P}=k+k_1$, with
\begin{eqnarray}
k&=& O(1/s){\cal P}+O(1/s){\cal P}'+\vb\nonumber\\
k_1&=&(x_1-1) {\cal P}+O(1/s){\cal P}'+\vone -\vb
\label{kkeqn}
\end{eqnarray}
We want to calculate the transverse parts of the gluon momenta
$k_1$ and $k$ w.r.t. the
${\cal P}_1$ direction. The momentum of the first gluon $k$ is purely
transverse to order $1/s$ and does not directly enter into Higgs
production. To order $1/s$ its transverse part remains unchanged.
The momentum $k_1$ has a non-negligible component in the ${\cal P}$
direction. With respect to the ${\cal P}_1={\cal P} +k+k_1$
direction it becomes
\beqn
k_1&=&({x_1-1\over x_1}) ({\cal P}+k+k_1) +{\tilde \k}_1 + O(1/s)
\label{k1eff}
\eeqn
where the vector ${\tilde \k}_1$ is now transverse
to ${\cal P}_1$.
Solving this equation for $k_1$ we get
\begin{equation}
{k_1\over x_1}=(x_1-1){\cal P} +O(1/s){\cal P}'+{x_1-1\over x_1}\vb+{\tilde
\k}_1
\end{equation}
which by comparing to Eq.~(\ref{kkeqn}) gives:
\begin{equation}
{\tilde \k}_1={\vone -\vb\over x_1}-{x_1-1\over x_1}\vb={\vone \over x_1}-\vb
\end{equation}
Hence, the transverse momenta which must enter the form factor are $\k$ and
$\tilde\k_1$. We need to worry a little further as the coefficients of
the dot products can also be $x_1$ dependent.

To check this, and to check how $x_1$ enters into the diagram,
we consider the simplest case: the form factor ${\cal E}_1$ in the
case of elastic of pion scattering. way
is to consider ${\cal E}_1$ in the case of pion exchange.

First solve for the pion kinematics:
let $\Delta=(x_1-1)\ \p +2\beta_\Delta/s\ \p'+{\bf\Delta}$, with
$\p.\p=0$, $\p '.\p '=0$ and $\p .\p'=s/2$. The on-shell
condition for the final pion, $(\p +\Delta ).(\p +\Delta )=0$,
gives:
\beq\beta_\Delta =  - {{\bf\Delta}^2\over 2\ x_1}\eeq
and the total momentum transfer squared is:
\beq t={{\bf\Delta}^2\over x_1}=x_1(\tilde \k _1+\tilde\k )\eeq

As we are calculating the imaginary part of the diagram, we have a few
on-shell conditions for some of the quarks:
one of the incoming quarks is on-shell, call its momentum $r$, with
$r=x\ \p+2\beta_r/s\ \p '+\rrr $. The on-shell condition $r.r=0$ gives
\beq\beta_r= -{ \rrr ^2\over 2\ x}\eeq
The off-shellness of the other quark is:
\beq m^2=(\p -r).(\p -r)= {\rrr ^2\over x}\eeq
The momentum of the first gluon is given by
$k=2\alpha_k/s\ \p +2\beta_k/s\ \p '+\k $, and the last on-shell condition,
$(\p -r+k).(\p -r+k)=0$, gives:
\beq\beta_k = {\k ^2 \ x - 2\ \k .\rrr \ x +
\rrr ^2\over 2\ x\ (x - 1)}+O(1/s)\eeq
The second gluon is $k_1=\Delta-k$, with
\beq k_1\approx (x_1-1 )\ \p
 + { - 2\ x_1 \beta_k - x_1\ t
 \over  x_1\ s}\ \p ' - \k +{\bf\Delta}\eeq
Now, the dot product of the two quarks making up the proton is:
\beq A_2=
2 r.(p-r+\Delta)=-2(p-r+\Delta).(p-r+\Delta)
= {- m^2} x_1 + 2 \rrr .{\bf\Delta} - x\ t\eeq

The form factor emerges as a convolution of wavefunctions. These can only
depend on the dot products of the 2 quarks making the proton, hence on $A_1$
and $A_2$.
In the case where $x_1=1$, one gets:
\beq\int \psi^*({\rrr ^2\over x}) \psi({\rrr\over\sqrt{x}}
-\sqrt{x} {\bf\Delta})^2
d^2\rrr dx\eeq
which clearly is a function of $\bf\Delta$ only.
In the more general case, one gets:
\beq\int \psi^*({\rrr^2\over x}) \psi({r\over\sqrt{x/x_1}}-\sqrt{x/x_1}
{\bf\Delta})^2 d^2\rrr dx\eeq
So, for $t_1>>m^2$, one gets $F_1(t_1)$. However, this is not true in general,
and the form factor clearly has an extra $x_1$ dependence: $\bf\Delta=0$
does not lead to a form factor equal to 1. One can in fact make a rough
guess: the effective argument is
$\tilde t=t+<{m^2\over x}> (1-x_1)$ and $<{m^2\over x}>\approx 9\times 0.3^2
\approx 1$ GeV$^2$.

Hence we see that the ansatz is as follows: replace the transverse
vector $\k_1$ by $\tilde\k_1$, and multiply the overall argument by
$x_1$.  Note that this guarantees the infrared finiteness of the
answer when $k^2$ or $k_1^2={\tilde \k}^2/x_1\rightarrow 0$. Clearly,
this procedure straightforwardly generalises to the diffractive case
$pp\to X_1 X_2 H$.

\section{Higgs production}
Apart from the form factors $W_1$ and $W_2$, the preceeding formalism can be
applied to the production of any rare particle within a rapidity gap. We shall
now specialise the calculation to the production of the minimal standard model
Higgs
boson. We first calculate the effective coupling resulting from the triangle
graphs.

\subsection{ Higgs-gluon-gluon effective vertex}

\begin{figure}
\centerline{\psfig{figure=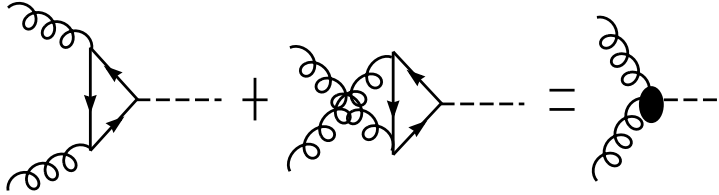,width=10cm}}
{\caption{\tenrm\baselineskip=12pt The Higgs-gluon-gluon vertex. All quarks
run around in the loop, but since the Higgs-quark-quark coupling is
proportional to its mass, the top quark contribution dominates.
}
\label{fig3}}
\end{figure}

By calculating the diagrams in Fig.~\ref{fig3} we arrive at the
following effective Higgs-gluon-gluon vertex:
\beq
\delta^{ab} (k_1\cdot k_2 g^{\mu\nu}-k_1^\mu k_2^\nu){W_1 \over m_H^2}
+ (k_1^2 k_2^2 g^{\mu\nu} + k_1^\mu k_2^\nu k_1 \cdot k_2
    - k_1^\mu k_1^\nu k_2^2 - k_2^\mu k_2^\nu k_1^2){W_2 \over m_H^4}
\label{effvertt}
\eeq
where
\beq
W_1 = { [\sqrt2 G_f]^{1/2} \alpha_s m_H^2 \over 3\pi}  N_1 \qquad
W_2 = { [\sqrt2 G_f]^{1/2} \alpha_s m_H^2 \over 3\pi}  N_2
\eeq
and
\beqn
N_1 &=& 3 \int^1_0 dx \int^{1-x}_0 dy {1-4xy \over D }\\
N_2 &=& {3 m_H^2 \over k_1\cdot k_2} \int^1_0 dx \int^{1-x}_0 dy {1-2x-2y+4xy
 \over D}\\
D &\equiv&
1-(2k_1\cdot k_2/m_t^2) xy +(k_1^2/m_t^2)(y^2-y)+(k_2^2/m_t^2)(x^2-x)
\label{n1n2}
\eeqn

Because of the kinematics of Higgs production in a rapidity gap,
$k_1^2, k_2^2<<2k_1 \cdot k_2\approx m_H^2$, and we can approximate
the integral expression by setting $k_1^2, k_2^2$ equal to zero:
\beqn
N_1^o &=& 3 \int^1_0 dx \int^{1-x}_0 dy {1-4xy \over D^o }\\
N_2^o &=& 6 \int^1_0 dx \int^{1-x}_0 dy {1-2x-2y+4xy \over D^o}\\
D^o &\equiv& 1-(m_H^2/m_t^2) xy -i\epsilon
\label{n1n2approx}
\eeqn
Note that with $k_1^2$ and $k_2^2$ set equal to zero it is the
$i\epsilon$ prescription coming from the top quark propagators in the loop
that defines the contour.

Since the off-shellness of the gluons is small compared to the Higgs
mass, and since the tensor structure (\ref{effvertt}) will always contract
$k_1$ with the upper side ($\mu$) of the diagram and $k_2$ with the lower
side ($\nu$),
the $W_2$ term in Eq.~(\ref{effvertt}) can be ignored,
unless of course $|N^o_2|^2$ turns out to be abnormally large compared
to $|N^o_1|^2$.  As we explain in more detail in the Appendix,
$|N^o_2|^2$ is always at least 30\% smaller than $|N^o_1|^2$ for Higgs
masses of 1 TeV or less.  Thus we ignore $W_2$ and the
above effective vertex can be derived from the following momentum
space Lagrangian
\beq
{\cal L_{\rm eff}} =  A^a_\mu A_{a\nu} H
               (k_1\cdot k_2 g^{\mu\nu}-k_1^\mu k_2^\nu){W_1 \over m_H^2}
\label{leffhaa}
\eeq
The term above, while
not gauge invariant, forms part of the following gauge invariant term.
\beq
{\cal L_{\rm eff}} = {W_1 \over 2m_H^2} G_{\mu\nu}^a G_{a\mu\nu} H
\label{leffhgg}
\eeq
As a check we use our effective vertex to
calculate the decay for $H\to gg$ and we obtain the well-known result
\beq
\Gamma(H\to gg) ={|W_1|^2 \over 2! \cdot 4 \pi m_H }
\label{htogg}
\eeq
where the last 2! is an identical final state particle phase space
symmetry factor.\footnote{Our result for the effective vertex differs from the
result quoted by Bialas and Landshoff by a factor 2, presumably the above
symmetry factor.}

The approximate form of $N_1$ in Eq.~(\ref{n1n2approx}) can be evaluated
in closed form. Defining $a\equiv m_H^2/m_t^2$ we have
\beqn
N_1^o &=& 3 \int^1_0 dx \int^{1-x}_0 dy {1-4xy \over 1-a xy-i\epsilon}
\nonumber\\
   &=& {6\over a} \left[
1 - (1-4/a) \left(\arctan[\sqrt{a/(4-a)}]\right)^2
			\right]
\qquad\quad {\rm if\ } a\equiv {m_H^2\over m_t^2} < 4
\label{n1approx} \\
   &=& {6\over a} \left[
1 + (1-4/a) \left(\arctanh[\sqrt{a/(a-4)}]\right)^2
			\right]
\qquad\quad {\rm if\ } a\equiv {m_H^2\over m_t^2} >4
\nonumber\\
\nonumber
\eeqn
where $\arctanh(x)=\ln|(x+1)/(x-1)| - i\pi/2$. At $a=4$, $N^o_1=3/2$, and
$N^o_1$ is continuous for the entire range of $a$.
Note the change in sign in front of the second term
between the two expressions.

\subsection{The Higgs production cross section}
 From Eq.~(\ref{cross}) we get:
\beqn \sigma = {1\over \shat^3 }{(2/9)^2\over 8}
{\alpha_s^4\over (2\pi)^5 } && \int  {dx_1 dy_2 \over
x_1 y_2} {d^2 \vone  d^2 \vtwo} d^2 \vb d^2\V \delta(H^2-M_H^2)
\tilde{\cal M} \nonumber\\
&& 3{\cal F}(k,k_1,K_1,K)~3{\cal F}(k,k_2,K_2,K)
\label{almostthefinalexpression}
\eeqn
with $k_1^2\approx (\vb+\vone )^2+(1-x_1)(\vb^2-\vone ^2)$,
$k_2^2\approx (\vb+\vtwo)^2+(1-y_2)(\vb^2-\vtwo^2)$, $k^2\approx \vb^2$,
and similar expressions for $K^2$, $K_1^2$, $K_2^2$,
and the Higgs momentum
\beq
H^2=(p_1+p_2-p-p')^2=\shat(1-x_1+{\vtwo^2\over \shat y_2}) (1-y_2+{\vone
^2\over
\shat x_1})+(\vone +\vtwo)^2
\label{higgsmom}
\eeq

The form factors $ 3{\cal F}(k,k_1,K_1,K)~3{\cal F}(k,k_2,K_2,K) $
correspond to calculating the inelastic cross section $pp\rightarrow
HX_1 X_2$.  To calculate $pp\rightarrow ppH$ these factors are
replaced by an expression written in terms of the Dirac elastic form factor
$F_1(t)$ given in Eq.~(\ref{F1})
\beq
(3 F_1(t_1) F_1(t_2) )^2 \qquad{\rm where}\quad t_1={\vone ^2/ x_1}
\quad {\rm and} \quad t_2={\vtwo^2/ y_2}
\eeq
In order to make explicit the flat $s$ behaviour we first
use Eq.~(\ref{finalcalm}) to substitute for $\tilde{\cal M}$
\beqn
\sigma={\shat\over 18 }{\alpha_s^4\over (2\pi)^5 }  && \int
{dx_1 dy_2 \over x_1 y_2} {d^2 \vone  d^2 \vtwo} d^2 \vb d^2\V
\delta(H^2-M_H^2)
{\cal A} W_1 W_1'^*/m_H^4 \\
&& {\cal F}(k,k_1,K_1,K){\cal F}(k,k_2,K_2,K)
D(k_1^2)D(k_2^2)D(k^2)D(K_1^2)D(K_2^2)D(K^2) \nonumber
\eeqn
$\tilde{\cal M}$ brings in a factor of $\shat^4$ and we still
need a factor of $1/\shat$ so we solve the Higgs momentum $\delta$ function
for $x_1$ or $y_2$. (Note that scaling $\vone $ and $\vtwo$ by $\sqrt{s}$
produces
the wrong power of $\shat$.)  In order to facilitate presenting
the solution, we
linearise the kinematics in terms of $1-y_2$.  Hence
\beqn
 H^2 &\approx& \shat((1-x_1)+{\vtwo^2\over \shat} +(1-y)2)
{\vone ^2\over \shat})((1-y_2)+{\vone ^2\over \shat} +(1-x_1)
{\vone ^2\over \shat})+(\vone +\vtwo)^2\nonumber\\
&\approx& \shat((1-x_1)+{\vtwo^2\over \shat})((1-y_2)+{\vone ^2\over \shat})
+(\vone +\vtwo)^2 \\
\nonumber
\eeqn
Thus solving the $\delta$ function we get
\beq
1-y_2\equiv(1-y_2)={M_H^2-(\vone +\vtwo)^2\over \shat((1-x_1)+{\vtwo^2\over
\shat})}-{\vone ^2 \over \shat}
\eeq
This brings a
$1/\shat/((1-x_1)+\vtwo^2/\shat)$ from the $\delta$ function upon performing
the $dy_2$ integral and the answer behaves
like $\shat^0$.

There is an overall arbritrary angle in the integral with respect
to which we measure all angles. We choose to measure our angles
from $\vb$. Thus $d^2\vb = d\theta d\vb^2/2 = \pi d\vb^2$.
Thus we arrive at:

\begin{eqnarray}
\sigma&=&{\alpha_s^4\over 18 (2\pi)^4 } \int
dx_1~d\vb~{d^2 \vone ~ d^2 \vtwo~} d^2\V~
{1\over x_1 y_2
\left((1-x_1)+{\vtwo^2\over \shat}\right)}\nonumber\\
&\times &{\cal F}(k,k_1,K_1,K)\ {\cal F}(k,k_2,K_2,K) \\
&\times& D(k_1^2)\ D(k_2^2)\ D(k^2)\ D(K_1^2)\ D(K_2^2)\ D(K^2)\nonumber
\label{thefinalexpression}
\end{eqnarray}
As explained in section 2, we need to multiply the above expression by
a Regge factor, $\prod_{i=1,2}\left({s\over s_i}\right)^{\alpha(t_1)}$, with
$s_i=(p_i+H)^2$, and $t_i=(p-p_i)^2$. This is our final
expression for the cross section.  Note that although the Regge factor
is very important when one fits the purely soft cross sections (as in
section 2.3), it plays very little role here, because $M_H^2$, and
hence $s_i$ are sizeable fractions of $\hat s$.

\subsection{The Bialas-Landshoff Limit}
Bialas and Landshoff (BL)~\cite{BL} work out the leading term in
$(1-x_1)$, $(1-y_2)$ at $\vone =\vtwo=0$, and then reintroduce the
$\vone $, $\vtwo$ integration after reggeisation. In this limit, we
get: $k_1^2=x_1 \vb ^2$, $k_2^2=y_2 \vb ^2$, ${\cal A}= (8\ |\vb ^2|
W_1)^2 \shat^4$.  To compare more easily with BL, we do not eliminate
the on-shell Higgs condition, and write:
\beqn d\sigma&=&{4\over 81\times 8 \shat}
{\alpha_s^4\over (2\pi)^5 \shat^3 }{dx_1\over x_1}
{dy_2\over y_2} {d^2 \vone d^2 \vtwo}
\delta((1-x_1)(1-y_2) -{M_H^2\over \shat})\nonumber\\
&\times& [\int d^2 \vb  8\ |\vb ^2| D(\vb ^2)
			D(x_1 \vb ^2) D(y_2 \vb ^2) W_1]^2 \shat^4\nonumber\\
&=&{\alpha_s^4\over  81 \pi^5  }{dx_1\over
x_1} {dy_2\over y_2} {d^2 \vone  d^2 \vtwo}
\delta((1-x_1)(1-y_2) -{M_H^2\over \shat})\nonumber\\
&\times&[\int d^2 \vb \ |\vb ^2| D(\vb ^2) D(x_1 \vb ^2) D(y_2 \vb ^2)
W_1]^2
\label{bllimit}
\eeqn
which is identical to formula (4.2) in Ref.~\cite{BL} at
$\epsilon=\alpha'=\lambda=0$,
except for the factor of 2 which come from symmetrizing two identical particles
as discussed in section~4.1.

Note that this is not the only simplification in their work, as they also
assume that an exponentially falling propagator multiplying $F_1$ is a good
representation of the product of the form factors by the gluon propagators.
Furthermore, their estimate of the diffractive cross section comes from the
replacement of $F_1$ by 1, and their integrals then converge only because of
the Regge slope. We shall now see that, surprisingly, the improvements we have
introduced -- higher orders in $(1-x_1)$, full form factors,
propagators -- do not significantly modify the final answer.

\subsection{Results}
Eq.~(\ref{thefinalexpression}) is an eight-dimensional integral which
we evaluate with the Monte-Carlo program VEGAS~\cite{vegas}, both for
the elastic cross section, $pp\to ppH$, (corresponding to using the
form factors in Eqs.~(\ref{formone}) and~(\ref{formtwo})), and for the
diffractive cross section, $pp\to X\thinspace H$, (corresponding to
the form factors described in Eq.~(\ref{formdiff})).

\begin{figure}
\centerline{
\psfig{figure=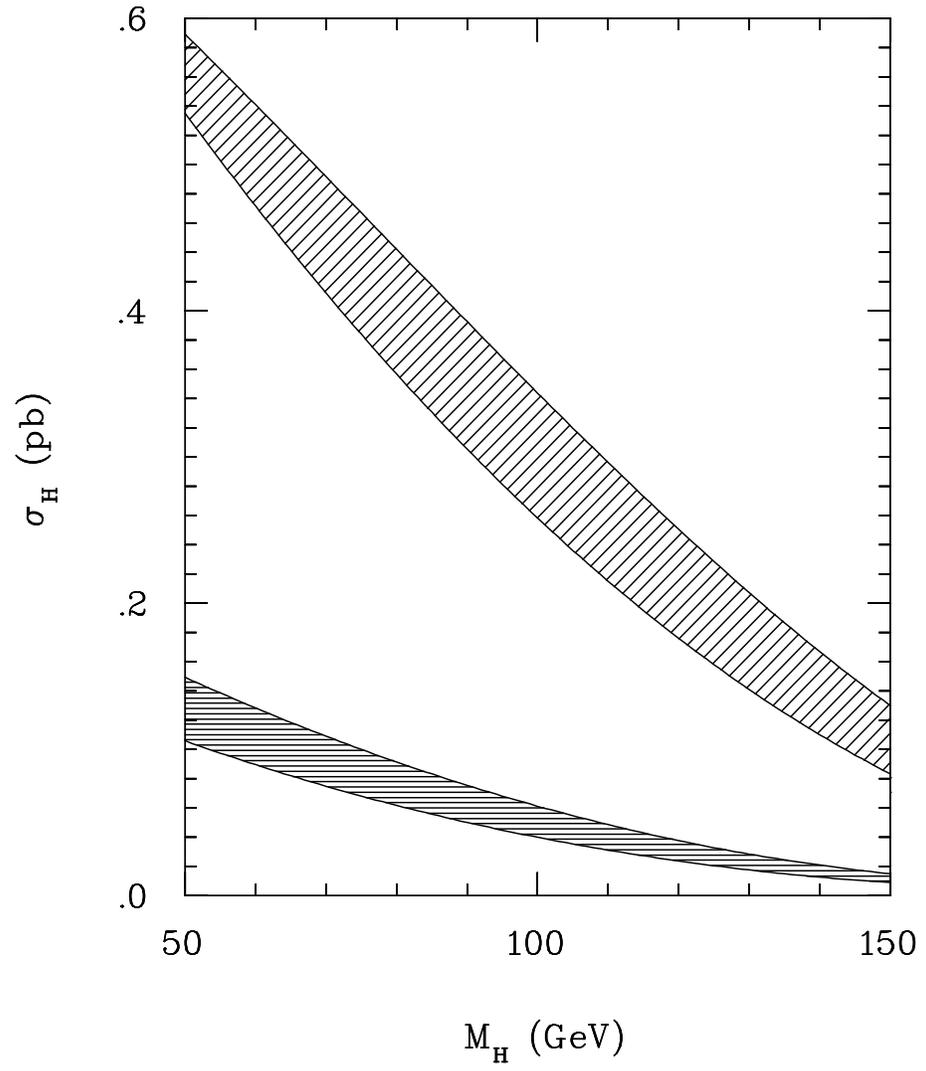,bbllx=3cm,bblly=1.5cm,bburx=16cm,bbury=22cm,width=10cm}}
{\caption{\tenrm\baselineskip=12pt The Higgs production cross section within a
gap at 1.8 Tev}
\label{fig1p8TeV}}
\end{figure}

\begin{figure}
\centerline{
\psfig{figure=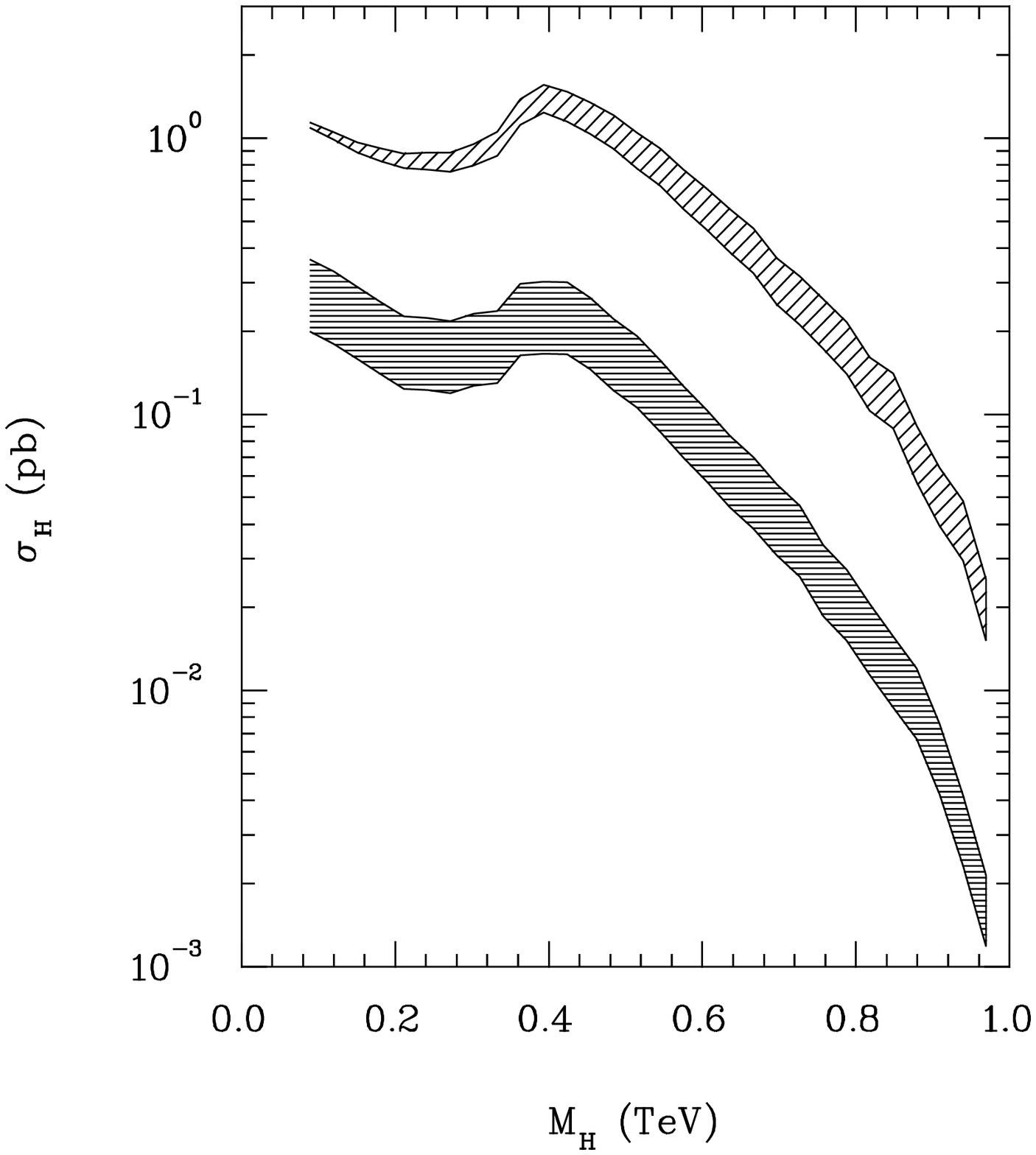,bbllx=3cm,bblly=1.5cm,bburx=16cm,bbury=22cm,width=10cm}}
{\caption{\tenrm\baselineskip=12pt The Higgs production cross section within a
gap at 10 TeV}
\label{fig10TeV}}
\end{figure}

\begin{figure}
\centerline{
\psfig{figure=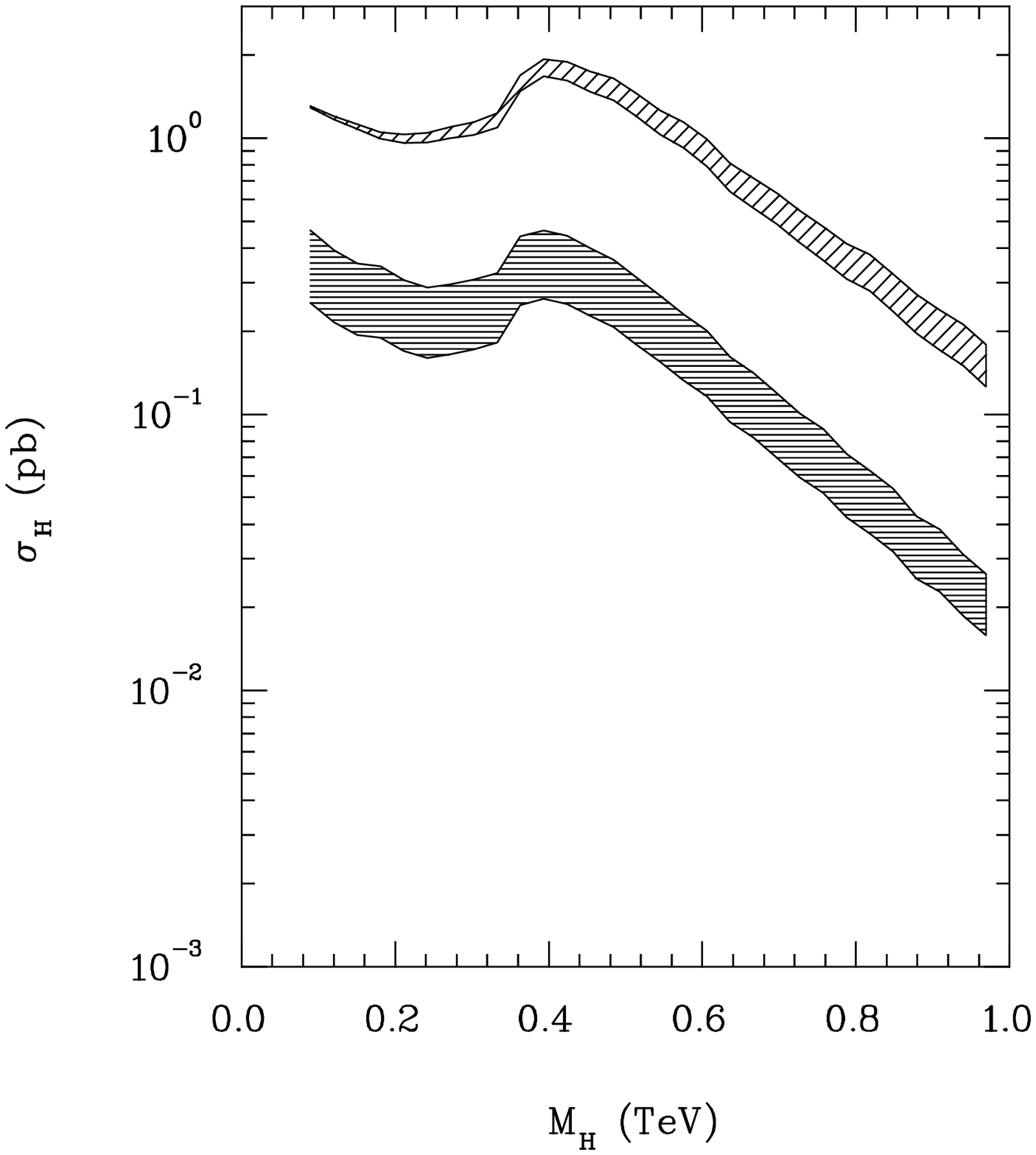,bbllx=3cm,bblly=1.5cm,bburx=16cm,bbury=22cm,width=10cm}}
{\caption{\tenrm\baselineskip=12pt The Higgs production cross section within a
gap at 14 TeV}
\label{fig14TeV}}
\end{figure}

We present our results for the Tevatron ($\sqrt{s}=1.8$ TeV), and the LHC
($\sqrt{s}=10$ TeV and
$\sqrt{s}=14$ TeV), in
Figs.~\ref{fig1p8TeV},\ref{fig10TeV},\ref{fig14TeV}. The
bands represent the effect of switching from a purely
perturbative propagator to a renormalisation group
improved propagator suppressed in the  infrared region. We
see that the extrapolation from the purely diffractive
cross sections of section~2 to Higgs production seems to
bring in relatively modest uncertainties of about a factor
2. The upper bands are for diffractive scattering, the
lower bands for elastic scattering.

It is striking to note that our calculation leads to essentially the
same conclusion as the Bialas-Landshoff estimate.  This can be
understood for the following reasons.  One phenomenologically
parametrises the model so as to correctly reproduce the soft elastic
and diffractive cross sections.  In the Higgs production diagrams, the
off-shellness of the gluons is rather small $\lsim 1$ GeV$^2$, and
comparable to that in the soft diffractive processes which we fit to.
The deep infrared region is cut off by the form factors.
Thus the behaviour of the gluon propagator in the UV region and the
deep IR region does not matter much, and the uncertainties arising
from the extrapolation to Higgs production are small.

The diffractive cross section (upper band) is on average about a
factor 5 higher than the elastic one (lower band).  We call attention
to the fact that it is really a factor 10 higher in the purely
perturbative case, and a factor 3 higher if we use the
constituent-gluon propagator. In other words the perturbative
calculation is the upper limit of the diffractive band and the lower
limit of the elastic band.  Given the smallness of the gluon
off-shellness, we consider the perturbative result less reliable.

We see that the cross section at the Tevatron is rather small, of the order of
a fraction of a picobarn, for light Higgs masses. It indeed gets rapidly cut
by the condition (\ref{massmax}) which precludes production of particles of
mass bigger than 200 GeV. If the Higgs is below 100 GeV, it will be
observable within a rapidity gap at the Tevatron. A handful of them may
already have been produced within the 60 pb$^{-1}$ available today.

\begin{figure}
\centerline{
\psfig{figure=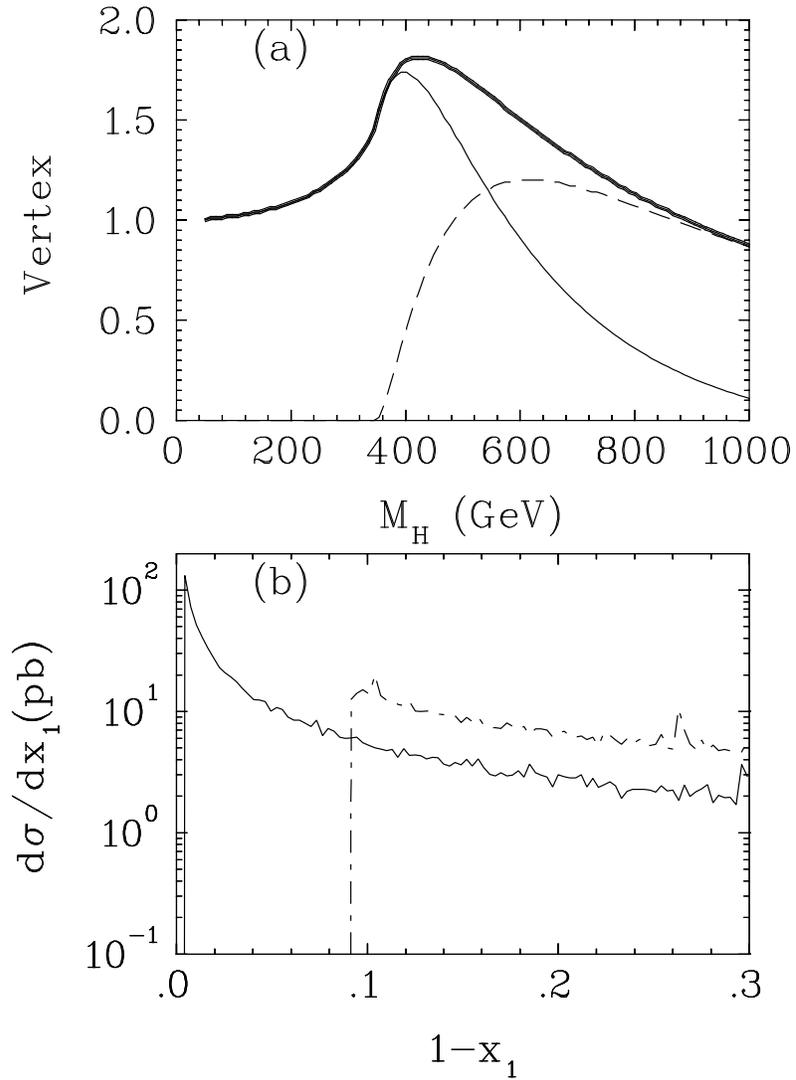,bbllx=3cm,bblly=1.5cm,bburx=19cm,bbury=25cm,width=10cm}}
{\caption{\tenrm\baselineskip=12pt
(a) The Higgs production vertex as defined by
{Eq.~(\protect\ref{n1approx})}.
The thin solid line shows the real part, the dashed line the imaginary part,
and the thick solid line the magnitude. (b) The $x_1$ distribution for a 100
GeV Higgs (solid line) and a 550 GeV Higgs (dashed line), for
$\protect\sqrt{s}=10$
TeV, in the diffractive case, using a constituent-gluon propagator.}
\label{fig6}}
\end{figure}

\begin{figure}
\centerline{
\psfig{figure=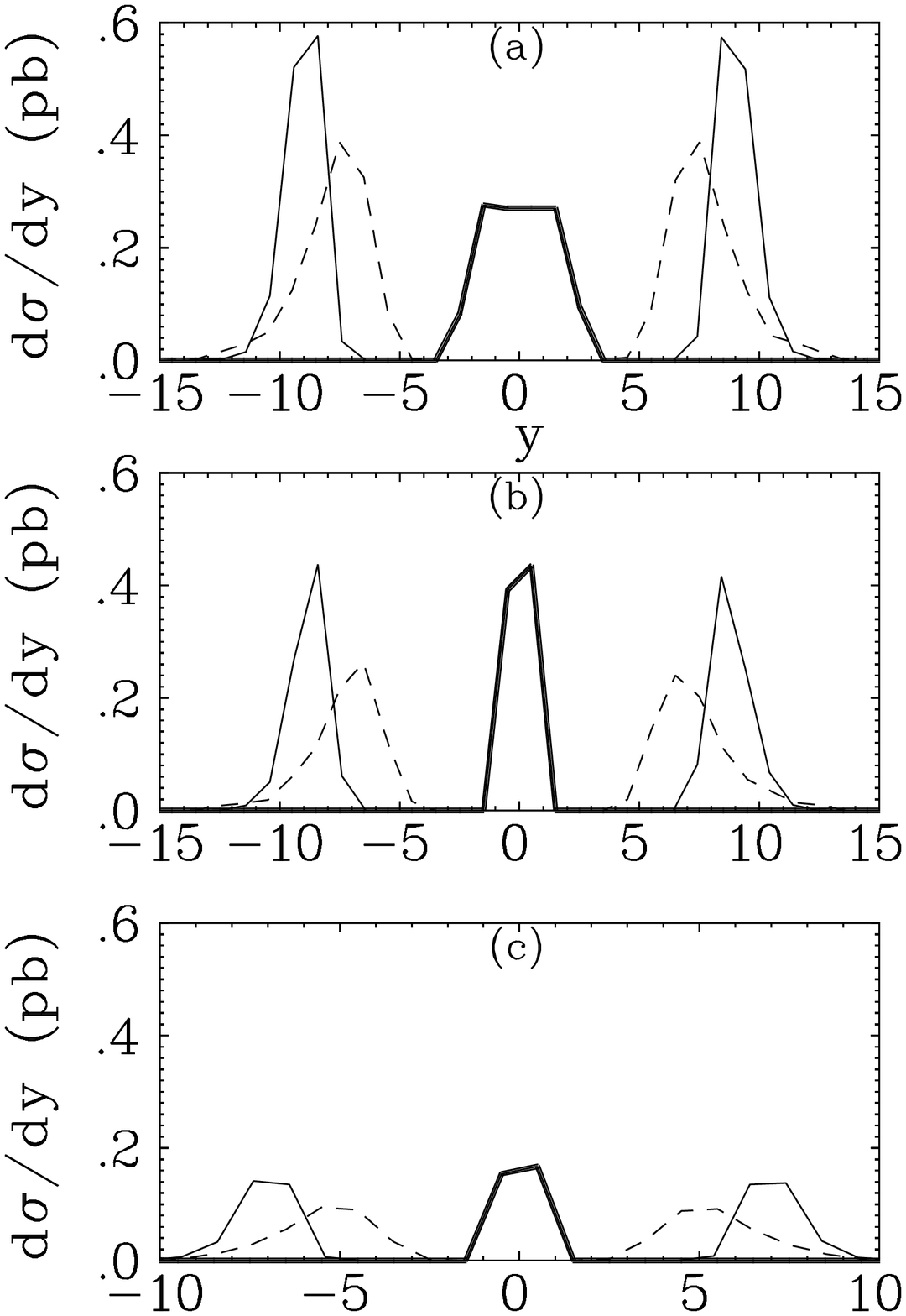,bbllx=3cm,bblly=1.5cm,bburx=19cm,bbury=25cm,width=10cm}}
{\caption{\tenrm The rapidity distribution of the Higgs (thick solid line),
of the partonic cluster (thin solid line), and decay pions (dashed line) at
(a) the LHC with $\protect\sqrt{s}=10$ TeV and a 100 GeV Higgs
(b) the LHC with $\protect\sqrt{s}=10$ TeV and a 550 GeV Higgs
(c) the Tevatron with a 100 GeV Higgs.
}
\label{figrap}}
\end{figure}

Fig.~\ref{fig6} helps us understand the shape of the Higgs production curves,
Figs.~\ref{fig1p8TeV},\ref{fig10TeV},\ref{fig14TeV}.
Fig.~\ref{fig6}a
shows the vertex as a function of the Higgs mass.  It rises with the
Higgs mass, until one reaches the threshold for top quark production (after
which the vertex is dominated by its imaginary part).
Fig.~\ref{fig6}b shows the $x_1$ distribution for a 100 GeV (solid line)
and a 550 GeV Higgs (dashed line), in the diffractive case, for
$\sqrt{s}=10$ TeV. Although the $x_1$ distribution is higher for the
higher mass (because the vertex is bigger), it gets cut off at
$x_1\approx 0.9$, after which one does not have enough partonic energy
to produce a 550 GeV Higgs.
Hence the low-mass Higgs cross section is higher because
it is dominated by large $x_1$. This also shows that
the dependence of the cross section on the cut-off $x_1>0.7$
(used to insure a rapidity gap) is very weak at low Higgs masses, and
gets progressively larger as the cross section decreases: the main cause
of its large fall-off at large Higgs masses is indeed the $x_1$ cut.
Putting these considerations together we can understand
the dip in the Higgs production curves before the rise peaking at about twice
the top mass.
Fig.~\ref{figrap}
gives an estimate of the size of the rapidity gap.  We
show the rapidity distribution of the Higgs (thick solid line) and
the partonic cluster (thin solid line).
As the latter will fragment into hadrons, we envisage a worst case
scenario, where the cluster decays into a forward proton and a
backward pion. The dashed curves give the pion rapidity distributions.
A gap is clearly present for $\sqrt{s}=10$ TeV, and the
Higgs peak becomes sharper as the mass of the Higgs increases. At the
Tevatron, we see that the gap is reduced to a few units of rapidity,
but remains present.
\section{Conclusion and further studies}
We have considered the lowest order QCD diagrams that give rise to
Higgs production within a rapidity gap. They correspond to single
pomeron exchange. We have found that the cross section is stable with
respect to the details of the infrared region, and that it is a
substantial fraction of the total Higgs production cross section
this way at the Tevatron, whereas the LHC will comfortably produce
1~pb of 600 GeV Higgses within a gap.

There are three main factors effecting the accuracy of our
predictions: the background, the survival of the gap, and higher order
corrections. The first two effects will reduce our estimate, the last
will increase it.

Although no conventional background is
expected, there will be direct heavy quark production in
the rapidity gap through the same color singlet exchange diagrams.
Some work along this line~\cite{heavyQbackground}
indicates that this background
will be tiny for Higgs masses greater than 200 GeV.
For Higgs masses less than 100 GeV the $c\bar c$ and $b \bar b$ background can
be a problem.

Another issue to consider is the survival of the rapidity gap.
Although the lowest order diagram produces a gap, further final-state
interactions between the out-going partonic clusters might destroy it.
Typical estimates for the survival of the gap give a number between 1
and 10\%~\cite{survival}. However, we want to point out that such an effect is
implicitly included in our calculation: we fit the soft diffractive
cross section, which already have to include such a survival factor,
and as the kinematics of the event is not very different from the soft
one, we expect those corrections to be small.

Finally, we have set up the calculation in such a way that higher
orders can be calculated \a la BFKL~\cite{BFKL}. In the elastic case
these corrections dramatically increase the cross section (e.g.~a
factor of $\sqrt{s}/M_H$). We expect a similar effect in our
calculation.

\section{ Appendix }
In this Appendix we will justify neglecting the $W_2$ term in
the effective vertex given in Eq.~(\ref{effvertt}) when calculating
Higgs boson production in a rapidity gap.
Given the cuts we've imposed to insure the survival the
gap, the approximate
form of $N_2$ given in Eq.~(\ref{n1n2approx}) can be used.
Defining $a\equiv m_H^2/m_t^2$ we have
\beqn
N_2^o &=& 6 \int^1_0 dx \int^{1-x}_0 dy {1-2x-2y+4xy \over 1-a xy-i\epsilon}\\
   &=& {-12\over a} \left[
5 - (1+4/a) \left(\arctan[\sqrt{a/(4-a)}]\right)^2 \right.\nonumber \\
&&\left.~~~~~~~~~~~
-{4\over \sqrt{a/(4-a)} } \arctan[\sqrt{a/(4-a)}]    \right]
 ~~~~~ {\rm if\ } a\equiv {m_H^2\over m_t^2} <4 \\
   &=& {-12\over a} \left[
5 +(1+4/a)\left(\arctanh\sqrt{a/(a-4)}
\right)^2\right.\nonumber \\
&&\left.~~~~~~~
-{4\over \sqrt{a/(a-4)} }\left( \arctanh\sqrt{a/(a-4)}\right)
\right]
{}~~~ {\rm if\ } a\equiv {m_H^2\over m_t^2} >4 \\
\nonumber
\label{n2approx}
\eeqn
At $a=4$, $N^o_2=-15+3\pi^2/2$.
It is the $i\epsilon$ prescription that tells us how to analytically continue
{}from $a<4$ to $a>4$ by replacing
\beqn
\sqrt{a\over 4-a} &\rightarrow& i \sqrt{a\over a-4} \nonumber \\
\arctan\sqrt{a\over 4-a} &\rightarrow&
i\left(\ln\left[{\sqrt{a/(a-4)}+1 \over \sqrt{a/ (a-4)}-1}\right]
						 - i{\pi\over2}\right)
		\equiv i \left( \arctanh\sqrt{a\over a-4}\right)
\label{anal}
\eeqn

\begin{figure}
\centerline{
\psfig{figure=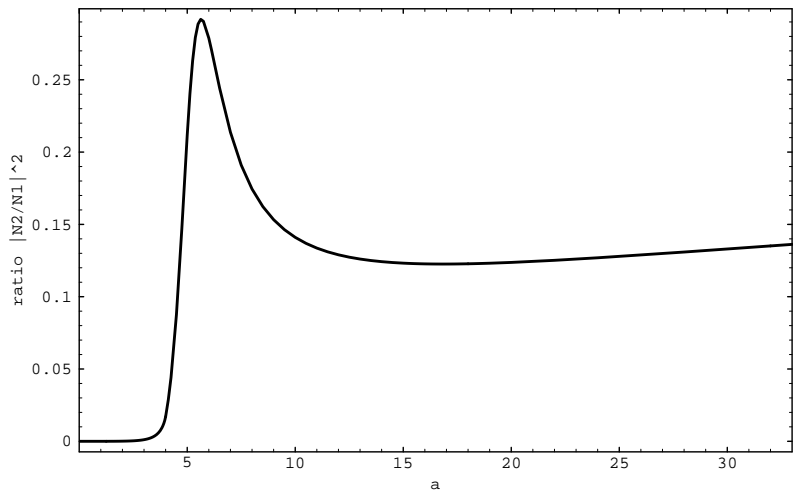,height=7cm}}
{\caption{\tenrm The ratio $|N^o_2|^2/|N^o_1|^2$ as a function of
$a\equiv m_H^2/m_t^2$ between $a=0$ and $a=33$. With a top
mass of $m_t=175$ GeV this corresponds to a Higgs mass $0<m_H<1$ TeV.}
\label{n2n1ratio}}
\end{figure}

We are interested in showing that the $N^o_2$ is never much greater
than the $N^o_1$ term. The analytic expression for $N^o_1$ is given in
Eq.~(\ref{n1approx}).  We plot the ratio $|N^o_2|^2/|N^o_1|^2$ in
Fig.~\ref{n2n1ratio}
and see that it is
always less than 0.30.
\setcounter{secnumdepth}{0}
\section{Acknowledgements}
JRC thanks A. Bialas, S. Brodsky
and D. Soper for their comments and suggestions.

\end{document}